\documentclass[prb,twocolumn,showpacs,preprintnumbers]{revtex4}
\usepackage{amsmath}
\usepackage{graphicx, epsfig}
\usepackage{dcolumn}

\def\be{\begin{equation}}
\def\ee{\end{equation}}
\def\bea{\begin{eqnarray}}
\def\eea{\end{eqnarray}}

\begin{document}

\title{ {\it Ab initio} estimate of temperature dependence of electrical
conductivity in a model amorphous material: hydrogenated amorphous silicon}
\author{T. A. Abtew}
\email{abtew@phy.ohiou.edu}
\author{Mingliang Zhang}
\email{zhang@phy.ohiou.edu}
\author{D. A. Drabold}
\email{drabold@ohio.edu}
\affiliation{Department of Physics and Astronomy, Ohio University, Athens, OH 45701}
\date{\today }

\begin{abstract}
We present an \textit{ab initio} calculation of the DC conductivity of
amorphous silicon and hydrogenated amorphous silicon. The Kubo-Greenwood
formula is used to obtain the DC conductivity, by thermal averaging over
extended dynamical simulation. Its application to disordered solids 
is discussed. The conductivity is computed for a wide range
of temperatures and doping is explored in a naive way by shifting the Fermi
level. We observed the Meyer-Neldel rule for the electrical conductivity
with $E_{{\small {\text{MNR}}}}$=0.06 eV and a temperature coefficient of
resistance close to experiment for \textit{a-Si:H}. In general,
experimental trends are reproduced by these calculations, and this suggests
the possible utility of the approach for modeling carrier transport in other
disordered systems.
\end{abstract}

\maketitle

\section{Introduction}

Amorphous semiconductors are among the most important electronic materials. 
\textit{a-Si:H} is the universal choice for TFT applications in laptop
displays and important photovoltaic material.   Thin films of a-Si:H are one of the promising
active elements for uncooled microbolometers\cite{fieque}  as employed in focal plane arrays for night vision applications. Here,  the temperature (T) dependence of the electrical
conductivity, and the Temperature Coefficient of Resistance (TCR) is of paramount importance. 
Experiments suggest that there are well-defined conductivity regimes:
variable-range hopping at the lowest temperatures, phonon-induced
delocalization at higher temperatures and ultimately a metallic form of
conduction (albeit with strong scattering). The venerable Meyer-Neldel rule%
\cite{MNR, yelon} (MNR) correlating the exponential prefactor of the DC
conductivity with the activation energy has been observed to be almost
universally valid in disordered systems.  A microscopic
understanding of these effects is highly incomplete, though significant
progress has been made on MNR\cite{yelon, overhof}.

Disordered solids pose definite challenges for transport modeling. The Boltzmann equation is best suited to long mean free paths, and is difficult to apply in these systems with localized states and defects. Moreover, the topological disorder must have a significant influence on the conduction and a realistic calculation of transport must include this disorder explicitly (eg through a credible structural model). 

Modern density functional simulations of materials routinely provide the
Kohn-Sham\cite{martin} eigenvalues and orbitals at each step in a thermal 
Molecular Dynamics (MD)
simulation. The natural approximate connection of these quantities to
electrical transport is provided by the Kubo-Greenwood formula (KGF)\cite%
{kubo,greenwood}. The first application of the KGF to an atomistic model
with computed electronic structure was by Allen and Broughton\cite{broughton}
for (metallic) liquid Si with a tight-binding Hamiltonian. In this paper, we
explore the utility of the KGF with Kohn-Sham spectral properties to
estimate the temperature dependence of the conductivity in \textit{a-Si} and 
\textit{a-Si:H}. An adiabatic approximation is used for modeling the
inelastic scattering of the phonons (by thermally averaging the KGF
over a constant T MD trajectory). The underlying idea is that if the thermal
fluctuations drive the unoccupied and occupied states close together (in the
spirit of Landau-Zener\cite{landau} tunneling), then the energy conserving $%
\delta$ functions are nonzero, and a finite contribution to the conductivity
is made by such instantaneous ``snapshots" if the dipole matrix element is
non-zero.  Doping, the MNR and TCR are obtained and shown to be in reasonable
agreement with measurements in \textit{a-Si:H}. The results are sufficiently
encouraging to justify a fuller exploration of the applicability of the
method to disordered systems more generally, including other amorphous
semiconductors and glasses and conducting polymers.

The paper is organized as follows. In Section II, we discuss the underlying
logic of the approach and some salient previous calculations, In Section
III, we briefly describe models and simulation procedures employed. In
Section IV, we present results for \textit{a-Si} and \textit{a-Si:H}
including the T-dependent conductivity, the effect of doping on the
conductivity and a discussion of MNR and TCR for \textit{a-Si:H}. Finally in
Section V we draw conclusions.

\section{Approach}

\subsection{Overview}

Previous work on amorphous semiconductors and glasses has shown that
Kohn-Sham eigenvalues conjugate to localized Kohn-Sham states are specially susceptible to the motion of the lattice\cite{91prl,atta} (thermal
fluctuation in the atomic coordinates involved in localized states leads to
a strong modulation of such eigenvalues and eigenvectors). Indeed, the RMS fluctuation of eigenvalues conjugate to localized eigenvectors is proportional to the localization of the eigenvector as measured
by inverse participation ratio\cite{atta}. 
Localized states appear in the gap, or in the spectral band tails. Since they are close to the Fermi level, they play a role in transport. The most elementary link between these states
and energies and the conductivity is the KGF, which is a natural approach to
strong scattering systems with short mean free paths. With this in mind, we
use the Kubo formalism and current \textit{ab initio} methods in conjunction
with fairly realistic (structurally plausible) supercell models of \textit{%
a-Si:H} to probe the temperature dependence of the DC conductivity. The
novelty of our work lies in 1) using Kohn-Sham states and energies in Kubo's expression
for the conductivity and 2) adopting a Born-Oppenheimer\cite{born} like
approximation of thermally averaging the KGF (using instantaneous atomic
configurations obtained in the course of a thermal simulation at constant
temperature as ``snapshots"). Such an approach certainly has limitations:
for low temperatures in which one expects variable range hopping between
defects the correct link to first principle simulation would seem to require
solutions of the time-dependent Kohn-Sham equations\cite{jundad,jundad1} or an approach more akin to a Miller-Abrahams\cite{miller} model of the conductivity. There are
several studies which use the KGF to compute the static lattice
conductivity of amorphous materials~\cite{holender,dong,ashwin}. The computed conductivity vanishes for localized states in a static lattice.

As an alternative to the weak scattering theory of Ziman\cite{ziman} for
liquid metals, investigators have used the KGF with variants of
Car-Parrinello\cite{martin} simulation to explore the conductivity of liquid
metals including for example, {\cal l}-NaSn alloys\cite{kaschner} and {\cal l}-Na\cite%
{silvestrelli}, with quite satisfactory results. It was found that Brillouin
zone sampling had to be treated with care, which is reasonable for a
metallic system. The success of such calculations is one motivation to
extend the approach to doped amorphous semiconductors\cite{dadboolbook}.
Another significant development of recent years is the Keldysh
Green's function approach offered to properly handle the effects of contacts
and finite potential drops for molecular electronics. An example of this
approach is the code TRANSIESTA\cite{trans}.

\subsection{Connection to many-body formulation}

The KGF has been applied on many occasions, and in liquids in a mode very similar to what we report here. However, there is usually no discussion of the underlying assumptions that connect the original many-body formulation of the KGF to its usual use with MD averages and single particle states. For example, the KGF as used in these computations is {\it not} formulated for use with inelastic processes, and thus the use of MD trajectories and averaging requires some discussion, which we offer in this section.

Consider a periodic perturbation $H_{int}$%
\begin{equation}
H_{int}=Fe^{-i\omega t}+F^{\ast}e^{i\omega t}   \label{per}
\end{equation}
acting on a system described via a many-body (electron and phonon) Hamiltonian $H$. The transition probability from initial state $\Psi_{i}$ to final
state $\Psi_{f}$ in the interval $d\nu_{f}$ under the action  of
an external field ($H_{int}$) is\cite{Landau77}%
\begin{equation}
\frac{2\pi}{\hbar}|\langle\Psi_{f}|F|\Psi_{i}\rangle|^{2}\delta(E_{f}-E_{i}-%
\hbar \omega)d\nu_{f}   \label{gol}
\end{equation}
where $\Psi_{f}$, $\Psi_{i}$ are eigenstates of many-body Hamiltonian $H$ 
\begin{equation}
H\Psi_{f}=E_{f}\Psi_{f},\text{ \ \ }H=H_{0}+H^{e-ph}.   \label{eax}
\end{equation}
Here, $E_{f}$ and $\Psi_{f}$ may be estimated by the ordinary time-independent
perturbation theory from the eigenfunctions $\Psi_{p}^{(0)}$ and eigenvalues 
$E_{p}^{(0)}$ of 
\begin{equation}
\text{ \ \ }H_{0}=H_{e}+H_{ion}.\text{\ }   \label{0}
\end{equation}

We adopt a  Born-Oppenheimer description 
\begin{equation}
H_{0}\Psi_{p}^{(0)}=E_{p}^{(0)}\Psi_{p}^{(0)}   \label{0s}
\end{equation}
 with,%
\begin{equation}
\Psi_{p}^{(0)}=\Phi_{e_{p}}\Theta_{v_{p}}   \label{BO}
\end{equation}
and $\Phi_{e_{p}}$ is a many electron wave function, $\Theta_{v_{p}}$ is a many
phonon wave function. The perturbation solution of (\ref{eax}) is%
\begin{equation}
\Psi_{i}=\sum_{q}a_{iq}\Psi_{q}^{(0)},\text{ \ \ }a_{iq}=%
\delta_{iq}+a_{iq}^{(1)}+a_{iq}^{(2)}+\cdots   \label{psi}
\end{equation}%
\begin{equation}
E_{i}=E_{i}^{(0)}+E_{i}^{(1)}+E_{i}^{(2)}+\cdots   \label{en}
\end{equation}
Detailed expressions can be found in standard textbooks. The coupling $%
H^{e-ph}$ between electrons and phonons need not be small, in principle
eigenfunctions and eigenvalues of $H$ can be calculated to any order of $%
H^{e-ph}$.

In $\omega \rightarrow 0$
limit, we only keep the interaction $F_{e}$ between electrons and external
field. Eq. \ref{gol} is modified to:
\begin{eqnarray}
&{}&\frac{2\pi }{\hbar }|\sum_{e_{p}e_{q}}[%
\sum_{v_{p}}a_{e_{f}v_{f};e_{p}v_{p}}^{\ast
}a_{e_{i}v_{i};e_{q}v_{p}}]\langle \Phi _{e_{p}}|F_{e}|\Phi _{e_{q}}\rangle
|^{2}  \nonumber \\
&\times &\delta \lbrack (E_{f}^{(0)}+E_{f}^{(1)}+E_{f}^{(2)}+\cdots ) 
\nonumber \\
&&-(E_{i}^{(0)}+E_{i}^{(1)}+E_{i}^{(1)}+\cdots )-\hbar \omega ]d\nu _{f}
\label{mp}
\end{eqnarray}%
Eq. \ref{mp} shows that for $\omega =0$, the conservation of energy is for the whole (electron+phonon) system. Phonons may assist the transition between two many-electron states with different energy if the phonons can serve as source or sink of energy.

At finite temperature, initial state $\Psi_{i}$ can be any one of all
possible states. To calculate the total absorption power in a sample, we
average initial state $\Psi_{i}$ with Boltzmann weighting, and sum
over all possible final states.

The Born-Oppenheimer approximation (BOA) starts with a stable lattice $A_{0}:$ ($X_{1}^{0},Y_{1}^{0},Z_{1}^{0},\cdots
X_{N}^{0},Y_{N}^{0},Z_{N}^{0}$) as its zero order configuration. The
sum over all possible phonon states $v_{p}$, one phonon, two-phonon, $%
\cdots $etc., in Eq. \ref{mp} means that we explore all possible configurations
of the lattice around our original conformation $A_{0}$. In the
configuration space spanned by all $N$ nuclear coordinates ($%
X_{1},Y_{1},Z_{1},\cdots X_{N},Y_{N},Z_{N}$), those configurations included in $\sum_{v_{p}}$ in Eq. \ref{mp} form a $3N$ dimensional region $\mathcal{D}$ around $A_{0}$.

The next step in the transition to current thermal simulations is the treatment of the ions as particles with trajectories governed by the classical equations of motion. For a system like a-Si:H, this is a defensible approximation, at least above the Debye temperature. Now, consider a MD simulation commencing from a configuration in the neighborhood of $A_0$.  One would then approximate Eq. \ref{mp} by%
\begin{equation}
\lim_{n\rightarrow \infty }\frac{1}{n}\sum_{j=1}^{n}\frac{2\pi }{\hbar }%
|\langle \Phi _{e_{f}}^{(j)}|F_{e}|\Phi _{e_{i}}^{(j)}\rangle|^{2}\delta
(\varepsilon_{f}^{(j)(0)}-\varepsilon_{i}^{(j)(0)})d\nu _{f}  \label{MDp}
\end{equation}%
where $n$ is the number of MD\ steps.  The assumption here is that the MD trajectory faithfully reproduces the dynamical processes implicit in Eq. \ref{gol}. In each MD step, the Kohn-Sham scheme gives the fully dressed single-particle energy for the electron-phonon coupling of the last MD step. In a sense, each MD step can be interpreted as a sum of one specific series of Feynman graphs in set $\mathcal{D}$,  if we classify Feynman graphs according to the configuration space points in $\mathcal{D}$. 

For a chosen basis set, one can expand many-electron wave function $\Phi$ as a linear combination of
Slater determinants of single particle functions $\psi_{r}$,%
\begin{equation}
\Phi_{e_{q}}=\sum_{q_{1}q_{2}\cdots q_{N}}e_{q_{1}q_{2}\cdots
q_{N}}c_{q_{1}}^{\dagger}c_{q_{2}}^{\dagger}\cdots c_{q_{N}}^{\dagger}|000\cdots0\rangle .
\label{mye}
\end{equation}
The interaction between electrons and external field ${\bf E}$ along the $x$ direction can be written in
single particle form as:
\begin{equation}
F_{e}=(\frac{eE}{2})\sum_{rs}\langle \psi_{r}|x|\psi_{s}\rangle
c_{r}^{\dagger}c_{s}   \label{op}
\end{equation}
from which the usual Kubo-Greenwood formula is obtained as we discuss in the next
subsection.

Depending upon the order of perturbation of Eq. \ref{psi}, Eq. \ref{mp}
describes various many-phonon assisted processes. This agrees with the
Green function formulation. Green function method\cite{Emin74,Emin77} starts from free single electron propagator and free single phonon propagator, this is contrast with Kubo formula which start with exact many-body wave function for the whole electron+phonon system. Technically, the former will be much easier than the later. Both acoustic and optical phonon assisted hopping\cite{Emin77} are worked out and applied to the transition rate from one localized state to another localized state, and the Meyer-Neldel relation is viewed as a consequence of assisted activation whenever large activation energies compared to typical excitation are involved \cite{Yelon90,yelon}.

\subsection{Application of KGF to disordered solids at finite temperature}

The derivation of the KG formula from linear response theory and the
fluctuation-dissipation theorem is available in the original literature\cite%
{kubo, greenwood}, and elementary derivations are provided in standard books
on transport in amorphous systems\cite{davis,overhof}. In these latter
derivations, first-order time-dependent perturbation theory (Fermi's Golden
Rule) is employed to deduce an expression for the AC
conductivity $\sigma(\omega)$. From either derivation, one expects the KGF
to be valid in the weak-field limit and for elastic scattering processes.
The form for the diagonal elements of the conductivity tensor for the static
lattice that emerges from the preceding discussion is: 
\begin{eqnarray}  \label{eqn1}
\sigma_{\alpha \alpha}(\omega)&=& \frac{2 \pi e^2\hbar}{\Omega m^2}%
\sum_{ni}|\langle \psi_n|p_\alpha |\psi_i\rangle|^2\frac{f_F(%
\varepsilon_i)-f_F(\varepsilon_n)}{\hbar\omega}  \nonumber \\
&\times& \delta(\varepsilon_n-\varepsilon_i-\hbar\omega)
\end{eqnarray}
where $\alpha=x,y,z$, f$_F$ is the Fermi distribution, $e$ and $m$ are the
electronic charge and mass, $p_\alpha$ is a component of the momentum operator, $\psi_i$ and $%
\varepsilon_i$ are the eigenstates and eigenvalues; $\Omega$ the cell
volume. The AC conductivity is then $\sigma(\omega)=\frac{1}{3}%
\sum_{\alpha}\sigma_{\alpha \alpha}$. In the rest of this paper,
single-particle Kohn-Sham states and eigenvalues are used for the $%
\{\psi\},\{\varepsilon\}$.

In the DC-limit ($\omega \rightarrow$0 ) the conductivity
takes the form 
\begin{equation}  \label{eqadd}
\sigma(T) = -\frac{1}{3}\sum_\alpha \int_{-\infty}^{\infty} \sigma_{\alpha
\alpha}(\varepsilon) \frac{\partial f_F(\varepsilon)}{\partial\varepsilon}
d\varepsilon
\end{equation}
where: 
\begin{equation}  \label{eqn3}
\sigma_{\alpha \alpha}(\varepsilon)= \frac{2 \pi e^2\hbar}{\Omega m^2}%
\sum_{ni}| {\langle
\psi_n|p_\alpha|\psi_i\rangle|^2\delta(\varepsilon_n-\varepsilon)\delta(%
\varepsilon_i-\varepsilon)},
\end{equation}
We include thermally-induced electron state and energy fluctuations near the
gap, by averaging expressions such as the preceding over a thermal
simulation. The DC conductivity is computed from
trajectory averaged quantities such as: 
\begin{equation}  \label{eqnaver}
\overline{\sigma}_{\alpha \alpha}(\varepsilon)= \frac{2 \pi e^2\hbar}{\Omega
m^2}\sum_{ni}| \overline{ {\langle
\psi_n^t|p_\alpha|\psi_i^t\rangle|^2\delta(\varepsilon_n^t-\varepsilon)%
\delta(\varepsilon_i^t-\varepsilon)}},
\end{equation}
where the bar denotes the average and we emphasize the dependence of the
various terms on the simulation time $t$. This average then picks up thermal
broadening effects in the density of states, and also include
time-dependence in the dipole matrix element. 
A Gaussian approximant for the $\delta$ function with a width of 0.05 eV is used
in our calculations. We have repeated our calculations for widths between
[0.01-0.1] eV,  and our results do not change appreciably. 
We insert thermally averaged diagonal elements as in Eqn. \ref{eqnaver} into
Eqn \ref{eqadd} to obtain the conductivities reported in this paper.

The KGF includes three possible transitions: (i)
localized state to localized state; (ii) localized state to
extended state; (iii) extended state to extended state. For amorphous
semiconductors, the upper tail of valence band and the lower tail of
conduction band are localized. Below room temperature, carriers are mainly
distributed in localized states, and (i) is dominant conduction mechanism.
With increasing temperature, there are more holes in the upper valence tail,
which are available for the electrons in lower side of mobility edge to
transit. In addition, the activation energy is lowered by thermal
fluctuations. At higher
temperature, (ii) becomes more important
until there are enough carriers in extended states, and (iii) will
becomes the most effective conduction channel.

It is believed\cite{street} that the temperature range from \textit{ca}
300K-700K is dominated by ``phonon-induced delocalization"\cite{fenz}, in
which the phonons aid the hopping (though not to the extent of polaron
formation). Phonon-induced delocalization has been computed from the
time-dependent Kohn-Sham equation\cite{jundad1}. 
It was observed that if electron energies became close (because of phonon-induced
variation in energies), mixing between the nearly degenerate states could occur,
which results in delocalization\cite{jundad}. This leads to a monotonic and
irreversible decrease in electron localization. In the present
calculation, the localization fluctuates (see Fig. 2 of Ref. \onlinecite{jundad}). Thus, our calculation
ignores the explicit phonon-driven delocalization except in so far as this
effect is captured in the thermal averaging scheme.

There are sources of error in any calculation of this type, which we must
mention: (1) We use the Kohn-Sham states computed in the LDA as input into the KGF. (2) We
employ very small structural models (\textit{ca} 70 atoms); (3) Except for a
few test cases, we have used a minimal (single-zeta) basis. In fact we do
this in part because this approximation gives a gap close to the physical
gap in \textit{a-Si}; (4) Our simulations run for finite time, though we
have taken some care to verify that the reported conductivities are
adequately converged; (5) There are physical mechanisms that should broaden
the density of states. Two of these arise from
Brillouin zone integration and the sparse and discrete sampling of the tail
states due to the small cell size. (6) Classical dynamics (no quantization of the
vibrations) is employed throughout. These errors are largely systematic, and
as such, difficult to measure.

On the other hand, there are conditions that ameliorate the situation. Where
point (1) is concerned: quasiparticle calculations for crystalline Si
computed in the GW approximation and the Kohn-Sham states are nearly
identical\cite{hybertsen}. Also, there is a strong thermal averaging effect:
even at room temperature, localized tail states can fluctuate by tenths of
an eV (far greater than $kT$); this helps significantly in connection with
points (2), (3) and (5) above. Finally, there is a convincing body of
empirical evidence that the KGF employed as we do here (or with more
primitive Hamiltonians) provides reasonably quantitative predictions for the
DC conductivity. The use of a SZ basis is convenient computationally, and
has the helpful feature that the gap is much closer to experiment than a
complete basis calculation.

\section{Hamiltonian and structural models}

The \textit{ab initio} local orbital code SIESTA \cite{siesta} was used to
perform the density functional calculations. We used a local density
approximation for the exchange (LDA) using the Perdew and
Zunger expression~\cite{pz}. Norm conserving Troullier-Martins pseudopotentials~\cite%
{tm} factorized in the Kleinman-Bylander form~\cite{kb} were used. We used a
minimal single $\zeta$ basis set for both Si and H\cite{basis}.We solved the self consistent
Kohn-Sham equations by direct diagonalization of the Hamiltonian and a
conventional mixing scheme. We used the $\Gamma(k=0)$ point to sample the
Brillouin zone in all calculations. No scissors correction was used since
the SZ optical gap is close to experiment (a scissors shift is necessary for
a polarized basis).

We have used two different well relaxed models of $aSi_{64}$ (64 Si atoms)~%
\cite{model} and $aSi_{61}H_{10}$ (61 Si and 10 H atoms)~\cite{model2}
models. We have prepared these models for six different temperatures 200K,
300K, 500K, 700K, 1000K, and 1500K. In each cases we followed the following
procedures. The two models were annealed to a particular temperature for 1.5
ps which is followed by equilibration for another 1.5 ps. Once the models
are well equilibrated, we performed a constant temperature MD simulation for
another 500 steps to obtain an average DC conductivity for the respected
models at a given temperature.

\section{Results}

\subsection{Amorphous silicon: \textit{a-Si}}

\subsubsection{Conductivity}

We have studied the electrical properties at different temperature by using
the inverse participation ratio ({I}) which is defined as 
\begin{equation}
\mathbf{I}=\sum_{i=1}^N[q_i(\varepsilon)]^2
\end{equation}
where {\it N} is the total number of atoms and $q_i(\varepsilon)$ is the Mulliken
charge residing at an atomic site $i$ for an eigenstate with eigenvalue $%
\varepsilon$ with $\sum_{i=1}^N[q_i(\varepsilon)]=1$. {I} is unity for an
ideally localized state and $1/N$ for an extended state.

In Fig.~\ref{{I}-aSi} we show the instantaneous IPR as a function of energy
for six different temperatures. As the temperature increases from 200K to
1500K, the optical gap is reduced and eventually at higher temperature all
the states become extended with no energy gap in the density of states. The
gap closes both because of thermal fluctuations on the eigenvalues and as a harbinger of the transition to a metallic state at sufficiently high temperatures. We
emphasize that these are only snapshots: the instantaneous IPR of a
well-localized state can vary\cite{jundad} by a factor of $\approx 2$. This
is understood as a consequence of the fluctuation of the eigenvalues (as
eigenvalues approach, the localized states associated with the eigenvalues
tend to strongly mix; mixed states (involving more than one defect center)
exhibit a reduced localization)\cite{dong}. 
\begin{figure}[h]
\begin{center}
\includegraphics[angle=0, width=0.48\textwidth]{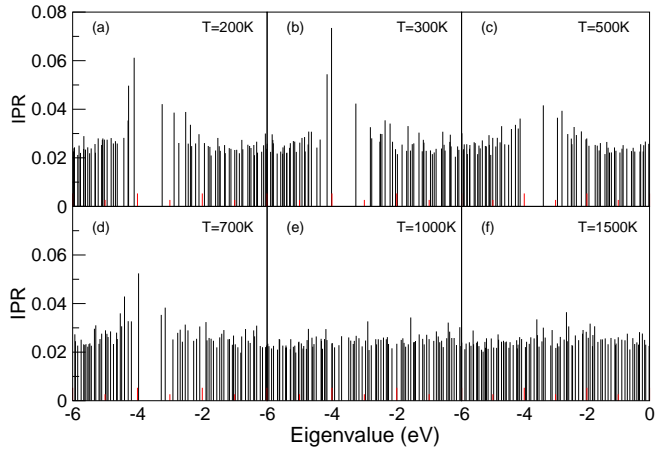}
\end{center}
\caption{Instantaneous snapshot of IPR (\textbf{I}) versus energy for
different temperatures in $aSi_{64}$ model. The Fermi level falls in the gap
near $E=-3.8eV$. Localization is significant near the band edges, and note
that the gap closes at higher temperature: at 1500K the cell has melted, at
1000K the gap has closed because of thermal fluctuations in the eigenvalues.}
\label{{I}-aSi}
\end{figure}

The DC conductivity of $aSi_{64}$ is accumulated over 500 instantaneous
configurations for different temperatures: T=200K, 300K , 500K, 700K, 1000K,
1500K, and 1800K. At a temperature of 1800K, the system is actually a liquid
with a diffusion coefficient of $D\sim1.6\times10^{-4}~\text{cm}^2 \text{s}%
^{-1}$ and a DC conductivity of $\sim0.3\times10^{4}~\Omega^{-1} \text{cm}%
^{-1}$ which is reasonably close to 
the measured value of $(1.0-1.3)\times10^{4}~%
\Omega^{-1} \text{cm}^{-1}$ \cite{glasov} and value of $1.75
\times10^{4}~\Omega^{-1} \text{cm}^{-1}$ obtained in another computation.\cite{car}.

\begin{figure}[h]
\begin{center}
\includegraphics[angle=0, width=0.48\textwidth]{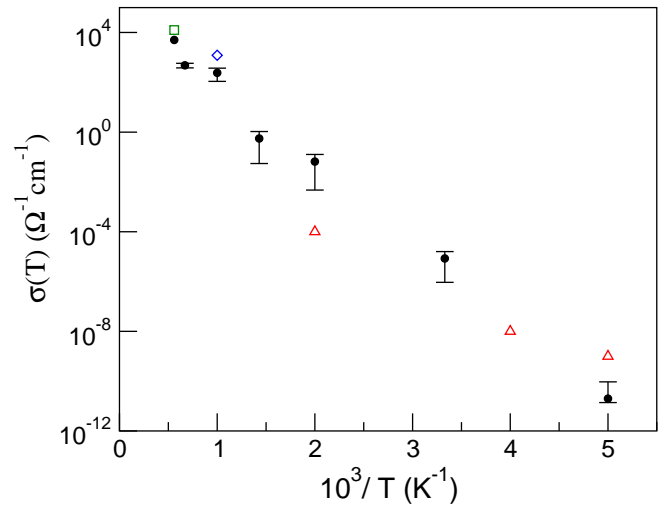}
\end{center}
\caption{(color online) DC conductivity of intrinsic $aSi_{64}$ averaged
over 500 configurations computed at different temperatures. The solid
symbols are from our work, open symbols are from experiment: open square
from Ref.~\onlinecite{glasov}, open diamond from Ref.~\onlinecite{ashwin}
, and open triangle from Ref.~\onlinecite{lewis}. The error bars are from
equilibration time and discrete sampling of the density of states.}
\label{cond1-aSi64}
\end{figure}

In Fig.~\ref{cond1-aSi64} we have shown the DC conductivity of $aSi_{64}$ as
a function of temperature. The results from experiment for selected
temperatures are also shown. As we can see from Fig.~\ref{{I}-aSi}, increase
in the temperature of the system enhances delocalization of the states and
eventually closes the optical gap to change the
material from semiconductor to metal. In doing so the DC conductivity
changes from $0.31\times10^{-10}~\Omega^{-1} \text{cm}^{-1}$ for T=200K to $%
0.24\times10^{3}~\Omega^{-1} \text{cm}^{-1}$ for T= 1000K.

The temperature dependence of DC conductivity can be written as 
\begin{equation}
\sigma = \sigma_oe^{(-E_a/k_BT)}
\end{equation}
where $E_a$ is the activation energy ($E_a=E_C-E_F$ or $E_a=E_F-E_V$) and $%
\sigma_o$ is the pre-exponential factor of the conductivity. By dividing the
dc conductivity in two regions of low temperature ($T < 450 K$) and high
temperature ($T > 450 K$) we extracted the $E_a$ and $\sigma_o$. For low T,
we have obtained $E_a\sim$ 0.34 eV and $\sigma_o \sim 4~\Omega^{-1} \text{cm}%
^{-1}$. For high T, $E_a\sim$ 0.45 eV and $\sigma_o \sim 1\times10^{4}
~\Omega^{-1} \text{cm}^{-1}$.

\subsection{Doping}

It is known that doping and temperature change result in a shift in the
position of Fermi level within the optical gap \cite{williams79,roedern79}.
In our simulation, we have computed the DC conductivity for a given doping
by shifting the Fermi level from its intrinsic position towards the
conduction band edge or valence band edge in steps of 0.1eV. This procedure allows us to ``scan" the
optical gap and compute conductivity for different doping levels of $n$-type
as well as $p$-type. Our scheme is quite different from experiment. For
example, it is known that the energy of defect states depends upon the location of the
Fermi level\cite{streetdope}, so that a proper calculation of doping (meaning with explicit substitution of the donor or acceptor species) should be carried out self-consistently.  Our procedure is a highly idealized version of the ``doping problem", which shows how the conductivity varies for without compensation effects. Because of these effects, it is not easy to correlate the conductivities we predict for a specific Fermi level position with the experimental concentration of dopant atoms.  We are presently extending this work by explicitly including B and P impurities in the system and computing the
conductivities separately for each doped model system.\cite{mingliang}

\begin{figure}[h]
\begin{center}
\includegraphics[angle=0, width=0.48\textwidth]{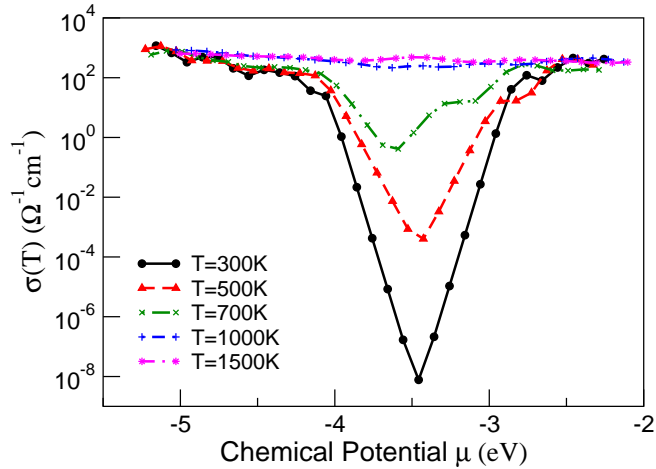}
\end{center}
\caption{(color online) DC conductivity of $aSi_{64}$ as a function of doping, averaged over
500 configurations computed at different temperatures versus chemical
potential.}
\label{cond2-aSi64}
\end{figure}

The computed DC conductivity for different temperatures as a function of
chemical potential is shown in Fig.~\ref{cond2-aSi64}. As the Fermi energy
shifts toward either the valence or conduction band from mid gap the DC
conductivity increases. At higher temperature, since the optical gap closes,
shifting the Fermi level (doping as $n$-type or $p$-type) doesn't yield any
significant change on the conductivity.

\subsection{Hydrogenated amorphous silicon: \textit{a-Si:H}}

\subsubsection{Conductivity}

In the same way we that analyzed $aSi_{64}$ in the previous section, we have
started our analysis of $aSi_{61}H_{10}$ by computing its electronic
properties from the inverse participation ratio. In Fig.~\ref%
{{I}-aSiH}, we have shown the IPR \textbf{I} of $aSi_{61}H_{10}$ for
different temperatures. As can be seen from the figure, the optical gap
decreases with increasing the temperature which is attributed to phonon
induced delocalization and structural rearrangements.

\begin{figure}[htpb]
\begin{center}
\includegraphics[angle=0, width=0.48\textwidth]{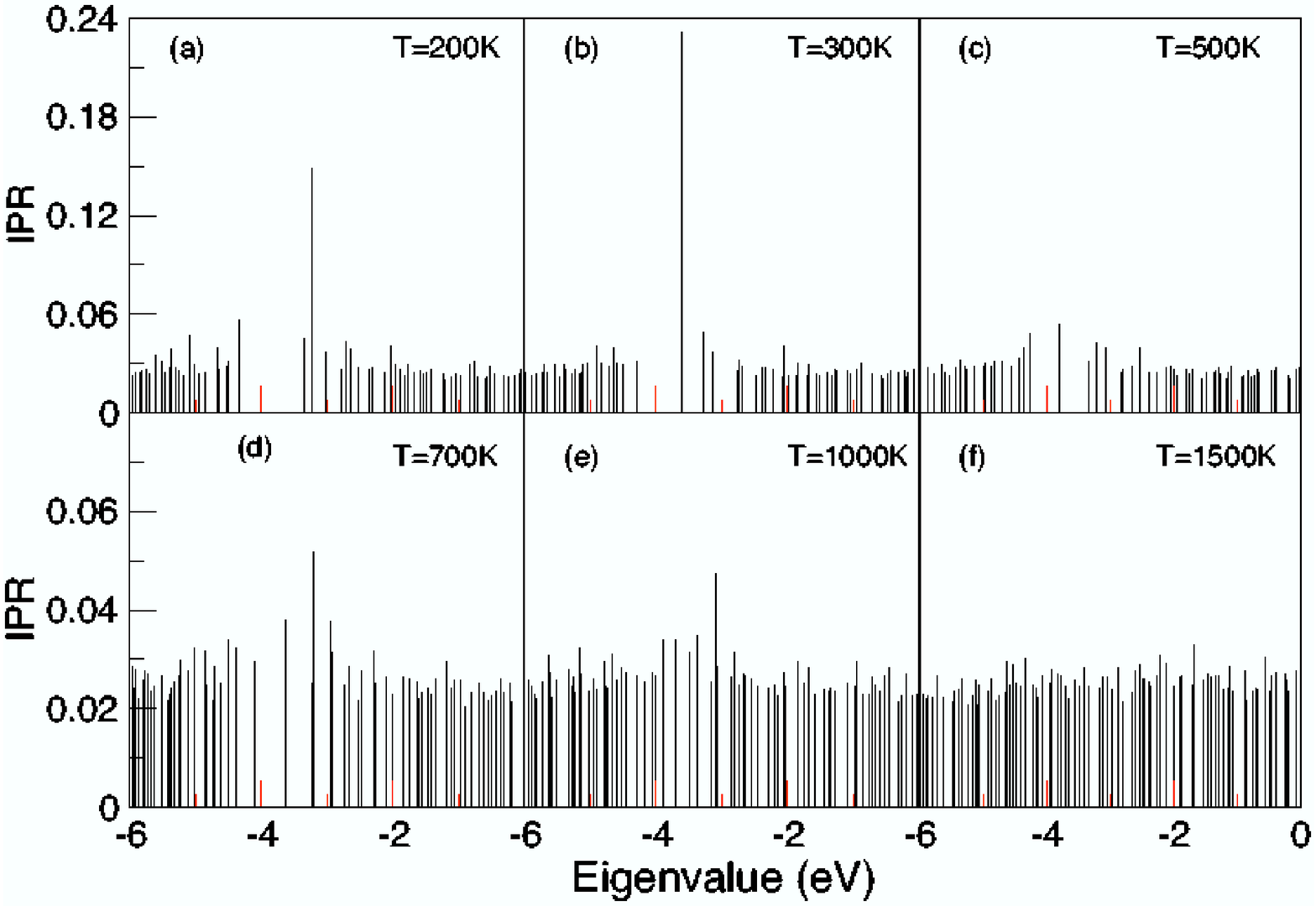}
\end{center}
\caption{IPR (\textbf{I}) versus energy for different temperatures in $%
aSi_{61}H_{10}$. Similar comments to Fig. 1}
\label{{I}-aSiH}
\end{figure}

\begin{figure}[htbp]
\begin{center}
\includegraphics[angle=0, width=0.48\textwidth]{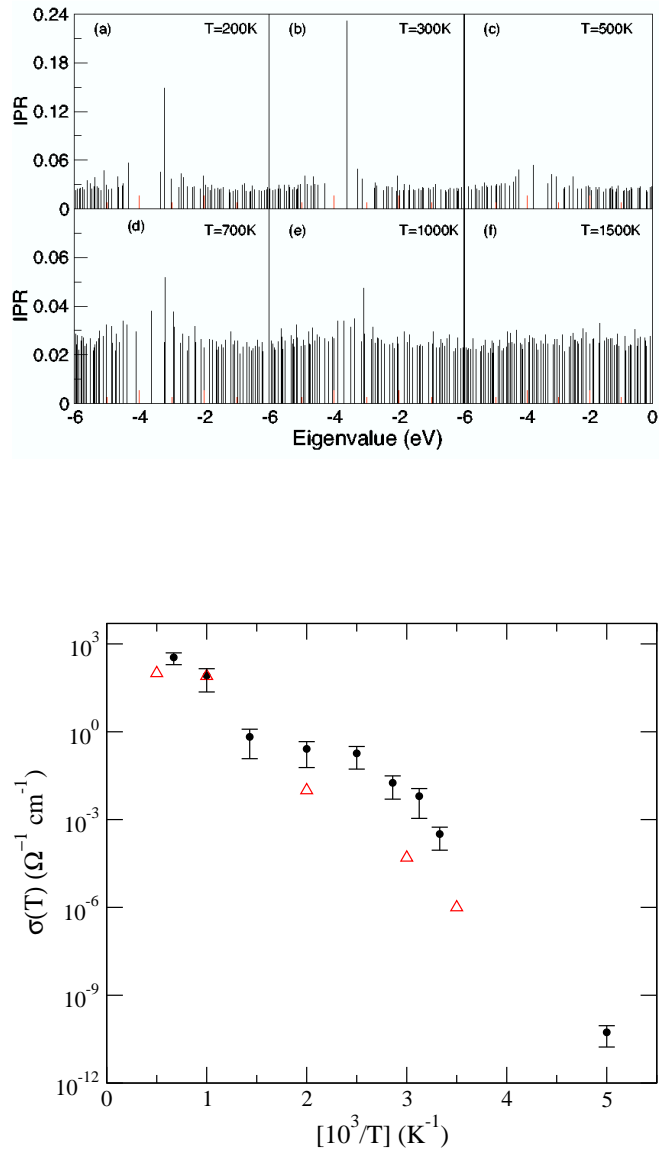}
\end{center}
\caption{(color online) DC conductivity of $aSi_{61}H_{10}$ averaged over
500 configurations computed at different temperatures. The solid symbols are
from our work, open symbols are from experiment Ref.~\onlinecite{beyer}. The
error bars are from equilibration time and discrete sampling of the density
of states.}
\label{cond1-aSiH71}
\end{figure}

The DC conductivity of $aSi_{61}H_{10}$ as a function of temperature is
shown in Fig.~\ref{cond1-aSiH71} with comparison from experimental results
from Beyer \textit{et al.}~\cite{beyer}. As we can see from Fig.~\ref%
{{I}-aSiH}, increase in the temperature of the system enhances
delocalization of the states and eventually eliminating the optical gap to
change the property of the material from semi-conductor to metal. In this
case, the DC conductivity changes from $0.54\times10^{-10}~\Omega^{-1} \text{%
cm}^{-1}$ for T=200K to $0.83\times10^{2}~\Omega^{-1} \text{cm}^{-1}$ for T=
1000K.

For low T, we have obtained $E_a\sim$ 0.31 eV and $\sigma_o\sim5~\Omega^{-1} 
\text{cm}^{-1}$. For high T, $E_a\sim$ 0.36 eV and $\sigma_o\sim
5\times10^{3} ~\Omega^{-1} \text{cm}^{-1}$. These results are in a
reasonable agreement with the experimental results of Kakalios \textit{et al.%
} \cite{kakalios86}. In studying doped \textit{a-Si:H}, Kakalios \textit{et al.}
showed that for low T, the $E_a$ ranges from 0.16 to 0.21 eV with $\sigma_o
\sim (5-10)~\Omega^{-1} \text{cm}^{-1}$.

\begin{figure}[htbp]
\begin{center}
\includegraphics[angle=0, width=0.48\textwidth]{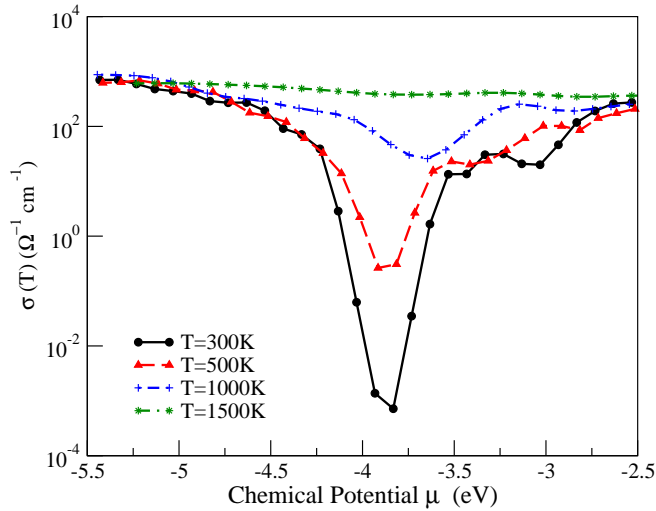}
\end{center}
\caption{(color online) DC conductivity of $aSi_{61}H_{10}$ as a function of 
doping,  averaged over
500 configurations computed at different temperatures versus chemical
potential. }
\label{cond2-aSiH71}
\end{figure}

The computed DC conductivity for different temperatures as a function of
chemical potential is shown in Fig.~\ref{cond2-aSiH71}. As in the case of $%
aSi_{64}$, the DC conductivity increases as the Fermi energy shifts toward
either the valence or conduction band from mid gap. At higher temperature,
since the optical gap is almost zero, shifting the Fermi level (doping as $n
$-type or $p$-type) doesn't yield any significant change on the conductivity.

As the temperature increases from 300K to 1000K, the contribution from the
matrix elements decreases while the contribution from the density of states
near the Fermi level increases. The temperature dependence
of the DC-conductivity is most affected by the density of states which is considerably
broadened for higher temperature.

\subsubsection{Meyer-Neldel rule}

In a band picture of conduction, one expects an exponential conductivity $%
\sigma=\sigma_0 exp(-E_a/kT)$. In disordered systems (not limited to a-Si:H)
there is a more complicated T-dependence, and in particular a ``kink" in the
conductivity in the vicinity of 400-500K\cite{peterthomasboolie}. The first
interpretation for this kink was the existence of distinct low-T and high-T
conduction mechanisms with significantly different activation energies
(slopes)\cite{lecomber}. It is argued that this interpretation is unlikely,
and that the effect arises from the temperature dependence of the Fermi
energy\cite{peterthomasboolie}.

For activated processes with activation energy greatly exceeding the
characteristic excitation energies available to the system, it is clear that
a fluctuation involving many small excitations will be needed to push the
system over the barrier. The larger the number of ways the necessary
fluctuation may be obtained the more likely it is for the process to occur. This 
line of thinking led to the concept of ``Multiple Excitation
Entropy" (MEE), which has clarified the origin of the Meyer-Neldel Rule (MNR) 
in a great many
different activated processes\cite{yelon}. One can think of the KGF as
representing a sum over pathways, in which case the number of available
paths or ``channels" is T-dependent and certainly increasing, reflecting an
entropic increase as discussed in MEE\cite{mov}.

\begin{figure}[htbp]
\begin{center}
\includegraphics[angle=0, width=0.48\textwidth]{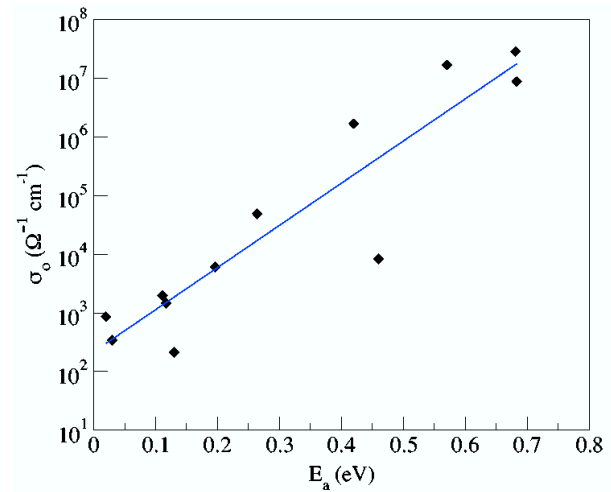}
\end{center}
\caption{(color online) The pre-exponential factor as a function of
activation energy (Meyer-Neldel rule) for $aSi_{61}H_{10}$. The dashed line
is an exponential fit which represent Eq.~(\protect\ref{mnr-eq}). It is
interesting to compare this result of simulation to a similar plot for a
large number of experimental samples as in Fig. 7.3 (Reference 
\onlinecite{street}). }
\label{meyer}
\end{figure}

For pre-exponential factor $\sigma_{0}$ and activation energy $E_a$, the MNR
may be expressed as: 
\begin{equation}  \label{mnr-eq}
\sigma_o=\sigma_{oo}e^{E_a/E_{\text{{\small {MNR}}}}}
\end{equation}
By performing a linear fit on the DC conductivity results, we identified the
intercept at $(1/T)=0$ to $\sigma_o$ and the slope to the activation energy $%
E_a$. There are number of experimental results on \textit{a-Si:H} which show
this exponential behavior with $E_{\text{{\small {MNR}}}}$=0.067 eV~\cite%
{carlsonM,fritzscheM}. By plotting $\sigma$ as a function of $1/T$ for
various dopants ($n$-type as well as $p$-type) we extracted $\sigma_o$ and $%
E_a$ for $aSi_{61}H_{10}$ and the results are shown in Fig.~\ref{meyer}. Our
calculation gives an exponential behavior of $\sigma_o$ as a function of $E_a$,
reflecting the Meyer-Neldel rule, with $E_{\text{{\small {MNR}}}}$= 0.060 eV.

The preceding shows that the mechanism(s) responsible for MNR are present in
our simulations. The enhancement of the conductivity with increased
activation energy may be qualitatively understood in our picture as being
due to the increase in electron-lattice coupling with increasing activation
energy (and therefore localization). Since localized states possess an
``amplified" electron-phonon coupling\cite{91prl,atta}, some compensation is
to be expected. In our picture, the MNR arises because the phonons treat
electrons with different localization (or $E_a$) differently, and the effect
runs in a direction consistent with experiment: for doping into the more
localized states, the electron-phonon coupling is larger and serves to
modulate the energies more strongly than for more weakly localized states
with smaller $E_a$.

\subsubsection{Temperature Coefficient of Resistance}

\begin{figure}[htbp]
\begin{center}
\includegraphics[angle=0, width=0.48\textwidth]{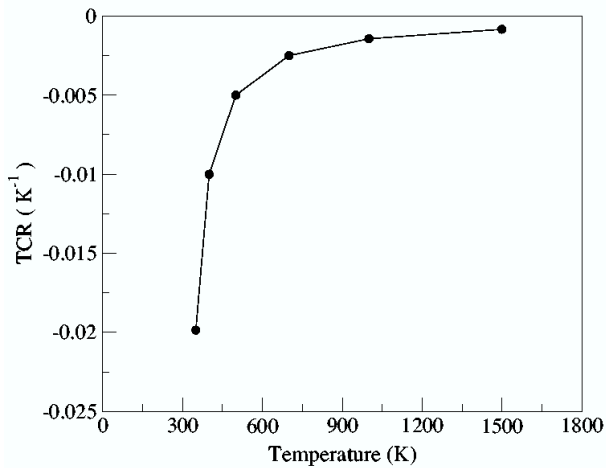}
\end{center}
\caption{The temperature coefficient of resistance (TCR) for $aSi_{61}H_{10}$
as a function of temperature.}
\label{TCR}
\end{figure}

The other fundamental characteristic of aSi:H is its high temperature
coefficient of resistance, which makes it a candidate for uncooled
microbolometer applications. The temperature coefficient of resistance (TCR)
is defined as 
\begin{equation}  \label{TCR-eqn}
TCR = \frac{1}{\rho_o}\frac{\rho-\rho_o}{T-T_o}
\end{equation}
where $\rho$ is a resistivity at any given temperature $T$ and $\rho_o$ is a
resistivity at a reference temperature $T_o$ (usually room temperature). The
computed result of TCR with a definition of Eq.~\ref{TCR-eqn} using $T_o=300K
$ for $aSi_{61}H_{10}$ is shown in Fig.~\ref{TCR}. The experimental TCR near room temperature is $-2.7\% K^{-1}$ for \textit{a-Si:H}~\cite%
{dutt}. Our calculations predict a TCR value of $\sim-2.0\% K^{-1}$ at $T=350K$ which
is in agreement with the experiment. Close to $T_0$ the value of TCR
is very sensitive to temperature and has a wide range of values $-(2.0-5.0)
\% K^{-1}$. 

\section{Conclusion}

We have presented a study of transport in an amorphous material. We used
Kubo-Greenwood formula for computing the DC conductivity of $aSi_{64}$ and $%
aSi_{61}H_{10}$ for different temperatures. We have also presented the
effect of doping on the DC conductivity. As the Fermi level approaches
either the conduction edge or valence edge we observe an increase in the DC
conductivity. Once the $E_f$ exceeds the ``mobility edge" we observe a weak
temperature dependence on the DC conductivity. Though it requires further
investigation (by using various dopants in $aSi_{64}$ and $aSi_{61}H_{10}$
), we observe the Meyer-Neldel rule with exponential behavior for the
pre-exponential factor $\sigma_{o}$. The computed result for TCR is in
good agreement with the experiment. Further study of this method
involves using a richer basis set, and fuller k-point sampling in the Brillouin
zone (or essentially equivalently, larger cells). 
Further ``deconstruction" of the KGF will be undertaken to explore the
instantaneous configurations that provide significant contributions to $%
\sigma$.

Our work shows that the simplest implementation of the KGF shows promise for
computing the electrical conductivity of disordered semiconductors, and this
success hints that it may be used with profit on other systems beside.

\section{Acknowledgements}

We thank the Army Research Office under MURI W91NF-06-2-0026, and the
National Science Foundation for support under grant No. DMR 0600073,
0605890. We acknowledge helpful conversations with J. David Cohen, A. Yelon,
B. Movaghar and P. Yue. Some of this work was carried out when DAD visited
the Institut de Ciencia de Materials de Barcelona with support from the
Programa de Movilidad de Investigadores of Minesterio de Educacion y Cultura
of Spain.


\begin{thebibliography}{99}
\bibitem{basis} This means that four (one) orbitals per site were used for Si (H).
\bibitem{model2} We removed three Si atoms from the $aSi_{64}$ model and
terminate all the dangling bonds except two with hydrogen. The newly formed
supercell is then relaxed well using conjugate gradient to form a $%
aSi_{61}H_{10}$ model (61 Si and 10 H atoms).
\end{thebibliography}
\end{document}